\documentclass[preprint,showpacs,preprintnumbers,amsmath,amssymb]{revtex4}

\usepackage{graphicx}
\usepackage{dcolumn}
\usepackage{bm}

\begin{document}

\preprint{}

\title{Decomposition of noncommutative $U(1)$ gauge potential}
\author{LIU Zi-Yu$^{1}$}
\author{LI Xi-Guo$^{2,3}$}
\email{xgl@impcas.ac.cn}

\affiliation{$^{1}$Department of Physics,Xianyang normal university,
Xianyang 712000, People's republic of China}
\affiliation{$^{2}$Institute of Modern Physics, Chinese Academy of
Science, P.O. Box 31, Lanzhou 730000, People's republic of China}
\affiliation{$^{3}$Center of Theoretical Nuclear Physics, National
Laboratory of Heavy Ion Collisions, Lanzhou 730000, China}

\begin{abstract}
We investigate the decomposition of noncommutative gauge potential
$\hat{A_{i}}$, and find it has inner structure, namely,
$\hat{A_{i}}$ can be decomposed in two parts $\hat{b_{i}}$ and
$\hat{a_{i}}$, here $\hat{b_{i}}$ satisfies gauge transformations
while $\hat{a_{i}}$ satisfies adjoint transformations, so dose the
Seiberg-Witten mapping of noncommutative $U(1)$ gauge potential. By
means of Seiberg-Witten mapping, we construct a mapping of unit
vector field between noncommutative space and ordinary space, and
find the noncommutative $U(1)$ gauge potential and its gauge field
tenser can be expressed in terms of the unit vector field. When the
unit vector field has non singularity point, noncommutative gauge
potential and gauge field tenser will equal to ordinary gauge
potential and gauge field tenser.

\end{abstract}

\maketitle

\section{Introduction}

The decomposition theory of the gauge potential in terms of unit
vector field was first assumed by Duan and Ge.\cite{du79}, then
subsequently by Cho\cite{ch80,ch81}. In 1999, Faddeev and Niemi
proposed a decomposition of the four dimensional $SU(N)$ Yang- Mills
field $A_{\mu}^{a}$\cite{fa99,fa9b}, recently, they also introduced
a novel, complete field decomposition in the Yang-Mills Lagrangian,
and suggested that due to fluctuations in the gauge invariant
condensate $\rho$ and the vector field $\mathbf{n}$, at short
distances both Newton¡¯s constant and the cosmological constant
become variable\cite{fa07}. The decomposition theory of the gauge
potential may provide a very powerful tool in establishing direct
relationship between differential geometry and topological
invariants in studding some topological properties of physical
systems. This has been effectively used to study the magnetic
monopole problem in $SU(2)$ gauge theory\cite{ho79,KI92,ho76}, the
topological structure of the Gauss-Bonnet-Chern theorem and Morse
theory\cite{du95}, the topological gauge theory of dislocation and
declinations in condensed matter physics\cite{du86,du92}, the
torsion structure on Riemann-Cartan manifold and knots\cite{li99},
the topological structure of Chern-Simons vortex\cite{li07}.

Recently the classical properties of noncommutative gauge theories
have been investigated in many directions\cite{gr00,gr01}. In fact
the Space-Time coordinates dose not commute was first assumed by
Snyder in 1947\cite{sn47}, and based on the Snyder construction, C.
N. Yang obtain another space-time that is compatible with the
Lorentz symmetry and also it is invariant under translations, this
space-time is discrete in the spatial coordinates\cite{ya47}. The
general noncommutative geometry was established by A. Conies, and
Yang-Mills theory on a noncommutative torus has been proposed as an
example\cite{co94}, for noncommutative manifolds $\mathbb{R}^{n}$,
the coordinates $x^{i}$ whose commutators are c-numbers,
\begin{equation}
[x^{i},x^{j}]=i\theta^{ij},
\end{equation}
with real $\theta$, where $\theta^{ij}$ is antisymmetric matrix,
i.e.
\begin{equation}
\theta^{ij}=-\theta^{ji},
\end{equation}
the product on noncommutative space is Moyal $\star$ product which
can be defined as\cite{ma49}
\begin{equation}
f\star
g=fg+\frac{i}{2}\theta^{ij}\partial_{i}f\partial_{j}g+\mathcal{O}(\theta^{2}).
\end{equation}

For noncommutative Yang-Mills theory, because of the $U(1)$
universality as noted by Gross and Nekrasov\cite{gr01}, it seems
that it is possible to everything about the $U(N)$ gauge theory can
be found in the $U(1)$ theory\cite{ba01}. In 1999, Seiberg and
Witten find an equivalence between ordinary gauge fields and
noncommutative gauge fields, which is realized by a change of
variables that can be described explicitly\cite{se99},
\begin{equation}
\hat{A_{i}}=A_{i}-\frac{1}{4}\theta^{kl}\{A_{k},\partial_{l}A_{i}+F_{li}\}+\mathcal{O}(\theta^{2}),
\end{equation}
\begin{equation}
\hat{\alpha}=\alpha+\frac{1}{4}\theta^{ij}\{\partial_{i}\alpha,A_{j}\}+\mathcal{O}(\theta^{2}),
\end{equation}
this is called Seiberg-Witten mapping, where the product on the
right hand side such as
$\{A_{k},\partial_{l}A_{i}\}=A_{k}\cdot\partial_{l}A_{i}+\partial_{l}A_{i}\cdot
A_{k} $ are ordinary matrix products, $\hat{A_{i}}$ is
noncommutative gauge potential, while $A_{i}$ is ordinary gauge
potential, $\hat{\alpha}$ is noncommutative gauge parameter,
$\alpha$ is ordinary gauge parameter, and so on.

In this paper, we will study the decomposition of noncommutative
$U(1)$ gauge potential by using Seiberg-Witten mapping. We decompose
the $U(1)$ Seiberg-Witten mapping in two parts, one part satisfies
gauge transformation, and the other satisfies an adjoint
transformation, and construct a mapping between unit vector field in
noncommutative space and in ordinary space, we discuss the inner
structure of noncommutative gauge potential and find the
noncommutative $U(1)$ gauge potential and its gauge field tenser can
be expressed in terms of the unit vector field. When the unit vector
field has non singularity point, noncommutative gauge potential and
gauge field tenser will equal to ordinary gauge potential and gauge
field strength.\\
\\
\\
\section{Decomposition of the Seiberg-Witten mapping of $U(1)$ gauge potential}
The gauge transformations of noncommutative Yang-Mills theory are
thus
\begin{equation}
\hat{A_{i}}\longrightarrow\hat{S}\star\hat{A_{i}}\star\hat{S}^{-1}+\partial_{i}\hat{S}\star\hat{S}^{-1},
\end{equation}
in which $\hat{A_{i}}$ is noncommutative gauge potential and
\begin{equation}
\hat{S}=exp(i\hat{\alpha})_{\star}
\end{equation}
is noncommutative gauge group, using Eq.(3) we can get
\begin{equation}
f\star (g+h)=f\star g+f\star h
\end{equation}

Suppose the noncommutative gauge potential $\hat{A_{i}}$ can be
written in two parts
\begin{equation}
\hat{A_{i}}=\hat{b_{i}}+\hat{a_{i}},
\end{equation}
sub Eq.(9) to Eq.(6), and using Eq.(8), we can get
\begin{equation}
\hat{A_{i}}\longrightarrow\hat{S}\star\hat{b_{i}}\star\hat{S}^{-1}+\hat{S}\star\hat{a_{i}}\star\hat{S}^{-1}+\partial_{i}\hat{S}\star\hat{S}^{-1},
\end{equation}
so can suppose $\hat{b_{i}}$ satisfies gauge transforation and
$\hat{a_{i}}$ satisfies an adjoint transformation
\begin{equation}
\hat{b_{i}}\longrightarrow\hat{S}\star\hat{b_{i}}\star\hat{S}^{-1}+\partial_{i}\hat{S}\star\hat{S}^{-1},
\end{equation}
\begin{equation}
\hat{a_{i}}\rightarrow\hat{S}\star\hat{a_{i}}\star\hat{S}^{-1}.
\end{equation}
the ordinary gauge potential can be written as\cite{du79}
\begin{equation}
A_{i}=b_{i}+a_{i},
\end{equation}
\begin{equation}
b_{i}\longrightarrow Sb_{i}S^{-1}+\partial_{i}SS^{-1},
\end{equation}
\begin{equation}
a_{i}\rightarrow S a_{i} S^{-1}.
\end{equation}
now we want find a mapping between $\hat{b_{i}}$ and $b_{i}$, and a
mapping between $\hat{a_{i}}$and $a_{i}$, this mapping should
satisfies Seiberg-Witten mapping, that is, for $U(1)$ group, those
mapping must satisfies
\begin{equation}
\hat{a_{i}}+\hat{b_{i}}=a_{i}+b_{i}+\frac{1}{2}\theta^{kl}(a_{l}+b_{l})(2\partial_{k}(a_{i}+b_{i})-\partial_{i}(a_{k}+b_{k})),
\end{equation}
like Seiberg-Witten mapping, we can suppose
\begin{equation}
\hat{a_{i}}=a_{i}+\frac{1}{2}\theta^{kl}f(A_{i}),
\end{equation}
\begin{equation}
\hat{b_{i}}=b_{i}+\frac{1}{2}\theta^{kl}g(A_{i}),
\end{equation}
where $f(A_{i})$ and $g(A_{i})$ is the function of $A_{i}$,
according to the principle of Seiberg-Witten mapping, the mapping of
$\hat{a_{i}}(A_{i})$ should satisfies
\begin{equation}
\hat{a_{i}}^{'}(A_{i})=\hat{a_{i}}(A_{i}^{'}),
\end{equation}
in which
\begin{equation}
\hat{a_{i}}^{'}=\hat{S}\star\hat{a_{i}}\star\hat{S}^{-1},
\end{equation}
and
\begin{equation}
A_{i}^{'}=S A_{i} S^{-1}+\partial_{i}SS^{-1}.
\end{equation}
 Equation (19) is solved by
\begin{equation}
\hat{a_{i}}=a_{i}+\theta^{kl}A_{l}\partial_{k}a_{i}-\frac{1}{2}\theta^{kl}a_{l}\partial_{i}a_{k},
\end{equation}
\begin{equation}
\hat{b_{i}}=b_{i}+\frac{1}{2}\theta^{kl}[A_{l}(2\partial_{k}b_{i}-\partial_{i}b_{k})-b_{l}\partial_{i}a_{k}].
\end{equation}
here we assume $b=0\Leftrightarrow \hat{b}=0$, so we decompose the
$U(1)$ Seiberg-Witten mapping in two parts, one part satisfies gauge
transformation (14), and the other satisfies an adjoint
transformation
(15).\\

\section{Seiberg-Witten mapping of unit vector field}
In order to decompose the noncommutative $U(1)$ gauge potential in
terms of unit vector field, we will construct an unit vector field
in commutative space, then establish a mapping between unit vector
field in noncommutative space and unit vector field in ordinary
space. Complex scalar field in noncommutative space is thus
\begin{equation}
\hat{\phi}=\hat{\phi_{1}}+i\hat{\phi_{2}},
\end{equation}
and complex scalar field in ordinary space is
\begin{equation}
\phi=\phi_{1}+i\phi_{2},
\end{equation}
using Seiberg-Witten mapping of complex scalar field\cite{ri03},
\begin{equation}
\hat{\phi}=\phi+\frac{1}{2}\theta^{ij}A_{j}\partial_{i}\phi,
\end{equation}
we can get
\begin{equation}
\hat{\phi_{1}}=\phi_{1}+\frac{1}{2}\theta^{ij}A_{j}\partial_{i}\phi_{1},
\end{equation}
\begin{equation}
\hat{\phi_{2}}=\phi_{2}+\frac{1}{2}\theta^{ij}A_{j}\partial_{i}\phi_{2},
\end{equation}
it is obvious their anti mapping is
\begin{equation}
\phi_{1}=\hat{\phi_{1}}-\frac{1}{2}\theta^{ij}\hat{A_{j}}\partial_{i}\hat{\phi_{1}},
\end{equation}
\begin{equation}
\phi_{2}=\hat{\phi_{2}}-\frac{1}{2}\theta^{ij}\hat{A_{j}}\partial_{i}\hat{\phi_{2}},
\end{equation}
we define a unit vector in noncommutative space as
\begin{equation}
\hat{n^{a}}=\frac{\hat{\phi^{a}}}{\sqrt{\hat{\rho}}},
\end{equation}
where
\begin{equation}
\hat{\rho}=\hat{\phi^{a}}\hat{\phi^{a}}
\end{equation}
it naturally has the constraint
\begin{equation}
\hat{n^{a}}\hat{n^{a}}=1.
\end{equation}
It is easy to prove that
\begin{equation}
\hat{n^{a}}=n^{a}+\frac{1}{2}\theta^{ij}A_{j}\partial_{i}n^{a},
\end{equation}
and its anti mapping
\begin{equation}
n^{a}=\hat{n^{a}}-\frac{1}{2}\theta^{ij}\hat{A_{j}}\partial_{i}\hat{n^{a}},
\end{equation}\\
\\

\section{Decomposition of noncommutative $U(1)$ gauge potential}
The decomposition of ordinary $U(1)$ gauge potential
is\cite{du79,li99},
\begin{equation}
A_{i}=\varepsilon_{ab}n^{a}\partial_{i}n^{b}-\varepsilon_{ab}n^{a}D_{i}n^{b},
\end{equation}
sub Eq.(35) into Eq.(36), we can obtain
\begin{eqnarray}
A_{i}=\varepsilon_{ab}\hat{n^{a}}\partial_{i}\hat{n^{b}}-\varepsilon_{ab}\hat{n^{a}}D_{i}\hat{n^{b}}
-\frac{1}{2}\theta^{kl}\varepsilon_{ab}[\hat{n^{a}}\partial_{i}\hat{A_{l}}\partial_{k}\hat{n^{b}}\nonumber\\
+\hat{A_{l}}\partial_{k}\hat{n^{a}}\partial_{i}\hat{n^{b}}-
\hat{n^{a}}D_{i}(\hat{A_{l}}\partial_{k}\hat{n^{b}})
-\hat{A_{l}}\partial_{k}\hat{n^{a}}D_{i}\hat{n^{b}}],
\end{eqnarray}
and the noncommutative gauge potential can be decomposed in follows
\begin{eqnarray}
\hat{A_{i}}=\varepsilon_{ab}\hat{n^{a}}\partial_{i}\hat{n^{b}}-\varepsilon_{ab}\hat{n^{a}}D_{i}\hat{n^{b}}
-\frac{1}{2}\theta^{kl}\varepsilon_{ab}[\hat{n^{a}}\partial_{i}\hat{A_{l}}\partial_{k}\hat{n^{b}}\nonumber\\
+\hat{A_{l}}\partial_{k}\hat{n^{a}}\partial_{i}\hat{n^{b}}-\hat{n^{a}}D_{i}(\hat{A_{l}}\partial_{k}\hat{n^{b}})
-\hat{A_{l}}\partial_{k}\hat{n^{a}}D_{i}\hat{n^{b}}]+\nonumber\\
\frac{1}{2}\theta^{kl}(\varepsilon_{ab}\hat{n^{a}}\partial_{l}\hat{n^{b}}-\varepsilon_{ab}\hat{n^{a}}D_{l}\hat{n^{b}})\cdot
[2\partial_{k}(\varepsilon_{cd}\hat{n^{c}}\partial_{i}\hat{n^{d}}\nonumber\\
-\varepsilon_{cd}\hat{n^{c}}D_{i}\hat{n^{d}})-
\partial_{i}(\varepsilon_{cd}\hat{n^{c}}\partial_{k}\hat{n^{d}}-\varepsilon_{cd}\hat{n^{c}}D_{k}\hat{n^{d}})],
\end{eqnarray}
when
\begin{equation}
Dn=0,
\end{equation}
using Eq.(36) we can get
\begin{equation}
A_{i//}=\varepsilon_{ab}n^{a}\partial_{i}n^{b},
\end{equation}
sub it to Seiberg-Witten mapping
\begin{equation}
\hat{A}_{i//}=\varepsilon_{ab}n^{a}\partial_{i}n^{b}+\frac{3}{2}\theta^{kl}\varepsilon_{ab}\varepsilon_{cd}
n^{a}\partial_{l}n^{b}\partial_{i}n^{d}\partial_{k}n^{c},
\end{equation}
because $a,b=1,2;c,d=1,2$,
\begin{equation}
\hat{A}_{i//}=A_{i//}[1+\frac{3}{4}\theta^{kl}\varepsilon_{ab}\partial_{k}n^{a}\partial_{l}n^{b}],
\end{equation}
using
\begin{equation}
\hat{F}_{ij}=\partial_{i}\hat{A_{j}}-\partial_{j}\hat{A_{i}}-i\hat{A_{i}}\star
\hat{A_{j}}+i\hat{A_{j}}\star\hat{A_{i}},
\end{equation}
we can obtain
\begin{equation}
\hat{F}_{ij//}=[1+\frac{3}{4}\theta^{kl}\varepsilon_{ab}\partial_{k}n^{a}\partial_{l}n^{b}]F_{ij//}
+\theta^{kl}\partial_{k}A_{i//}\partial_{l}A_{j//},
\end{equation}
where $F_{ij//}$ is ordinary gauge field strength,
\begin{eqnarray}
F_{ij//}=\partial_{i}A_{j//}-\partial_{j}A_{i//}\nonumber\\
=2\varepsilon_{ab}\partial_{i}n^{a}\partial_{j}n^{b},
\end{eqnarray}
so Eq.(44) can be rewritten as
\begin{equation}
\hat{F}_{ij//}=[1+\theta^{kl}\varepsilon_{ab}\partial_{k}n^{a}\partial_{l}n^{b}]F_{ij//},
\end{equation}
according to $\phi$-mapping theory\cite{du95}, we can obtain
\begin{equation}
\varepsilon_{ab}\partial_{k}n^{a}\partial_{l}n^{b}=\pi\delta(\overrightarrow{\phi})\varepsilon_{ab}\partial_{k}\phi^{a}\partial_{l}\phi^{b},
\end{equation}
and
\begin{equation}
\varepsilon_{ab}\partial_{k}\phi^{a}\partial_{l}\phi^{b}=\varepsilon_{kl\lambda_{1}\cdots\lambda_{m-2}}J^{\lambda_{1}\cdots\lambda_{m-2}}\left(
\frac{\overrightarrow{\phi}}{\overrightarrow{x}}\right),
\end{equation}
where $\overrightarrow{x}$ is vector in ordinary space, $m$ is the
dimension of space. Sub Eq.(47) and Eq.(48) to Eq.(42) we can get
\begin{equation}
\hat{A}_{i//}=A_{i//}[1+\frac{3}{4}\pi\theta^{kl}\delta(\overrightarrow{\phi})\varepsilon_{kl\lambda_{1}\cdots\lambda_{m-2}}J^{\lambda_{1}\cdots\lambda_{m-2}}\left(
\frac{\overrightarrow{\phi}}{\overrightarrow{x}}\right)],
\end{equation}
while
\begin{equation}
\hat{F}_{ij//}=F_{ij//}[1+\pi\theta^{kl}\delta(\overrightarrow{\phi})\varepsilon_{kl\lambda_{1}\cdots\lambda_{m-2}}J^{\lambda_{1}\cdots\lambda_{m-2}}\left(
\frac{\overrightarrow{\phi}}{\overrightarrow{x}}\right)],
\end{equation}
in which
$k,l,\lambda_{1},\cdot\cdot\cdot\lambda_{m-2}=1,2,\cdot\cdot\cdot
m$, from this equation we can see, when there is non singularity
point, noncommutative gauge potential and gauge field tenser will
equal to
ordinary gauge potential and gauge field strength.\\
\\
\section{Conclusions}
The gauge field in noncommutative space is different from ordinary
gauge field, this two fields can connected by the mapping which
supposed by Seiberg and Witten. The general decomposition theory of
gauge potential in ordinary space has been established by way of
geometric algebra, in which the gauge potential can be decomposed in
terms of unit vector field. In this paper, we prove the
noncommutative $U(1)$ gauge potential also has inner structure, and
constructed the decomposition of noncommutative $U(1)$ gauge
potential in terms of unit vector field by means of Seiberg-Witten
mapping, unlike the decomposition of ordinary gauge potential, there
are two kinds of unit vector field, the one is the unit vector field
in ordinary space, the other is the unit vector field in
noncommutative space, and we construct a mapping between unit vector
field in noncommutative space and in ordinary space. We also find
the Seiberg-Witten mapping can be decompose in two parts, one part
satisfies gauge transformation, and the other satisfies an adjoint
transformation. Finally, we calculate gauge potential and the gauge
field tenser and find if the unit vector field in ordinary space has
non singularity point, noncommutative gauge potential and gauge
field tenser will equal to ordinary gauge potential and gauge field
strength.
\section{acknowledgments}
This work was supported by the Talent introduction project of
Xianyang normal university.

\section{references}

\end{document}